\documentclass[runningheads,a4paper]{llncs}

\usepackage{amssymb}
\usepackage{graphicx}
\usepackage{tikz}
\usepackage{tabularx}
\usepackage{subcaption}
\usepackage{url}
\urldef{\mailsa}\path|pranav@barc.gov.in|
\urldef{\mailsb}\path|bose@barc.gov.in|
\newcommand{\keywords}[1]{\par\addvspace\baselineskip
\noindent\keywordname\enspace\ignorespaces#1}

\begin{document}

\mainmatter  

\title{On recent advances in 2D Constrained Delaunay triangulation algorithms}

\titlerunning{Recent advances in 2D CDT algorithms}

%
%
\author{Pranav Kant Gaur%
\and S. K. Bose%
}
\authorrunning{P.K. Gaur and S.K. Bose}

\institute{Computer Division, Bhabha Atomic Research Centre\\
Mumbai, India \\ \mailsa \\ \mailsb
}

%
%

\toctitle{Recent advances in 2D CDT algorithms}
\tocauthor{Pranav Kant Gaur}
\maketitle

\begin{abstract}
In this article, recent works on 2D Constrained Delaunay triangulation(CDT) algorithms have been reported. Since the review of CDT algorithms presented by de Floriani(Issues on Machine Vision, Springer Vienna, pg. 95--104, 1989), different algorithms for construction and applications of CDT have appeared in literature each concerned with different aspects of computation and different suitabilities. Therefore, objective of this work is to report an update over that review article considering contemporary prominent algorithms and generalizations for the problem of two dimensional Constrained Delaunay triangulation. 
\end{abstract}

\section{Introduction}
    Digital modelling and simulation of a natural phenomenon often requires discrete representation of the physical objects involved. Representation should be as close to the original object as possible and at the same time it must allow for a reasonably accurate simulation of the problem of interest. Delaunay triangulation satisfies later requirement\cite{shewchuk2002good} however, being a convex hull algorithm it does not necessarily preserve the object boundaries. Constrained Delaunay triangulation on the other hand relaxes the empty-circle criteria thereby making it possible to fulfil the former requirement. Consequently, not every element of the resulting triangulation is Delaunay but boundary constraints are preserved. As \cite{shewchuk2002constrained} points out, CDTs provide an advantage of object boundary preservation at the cost of strict compliance of each element of the mesh with the Delaunay property, however, in general, it also results in lesser number of additional points(called \textit{steiner points}) added to satisfy both Delaunay property for each element and preserving object boundary(a variant called \textit{Conforming constrained Delaunay triangulation}). \par
    Constrained Delaunay triangulation finds application in Path planning\cite{kallmann2005path}, Terrain modelling\cite{silveira2009optimization}, Geographic information systems\cite{qi2013computing}, PCB  design\cite{halama}, finite difference analysis\cite{mctaggart2004finite}, data-visualization\cite{Yang2009375} etc. CDT finds its application even in the field of parallel mesh generation as \cite{chew1997parallel} claims that using CDT as mesh generation approach results in reduction of communication cost and elimination of synchronization overheads as compared to other approaches used for mesh generation. \cite{devillers2003minimal} uses CDT to reconstruct a triangulation given its minimal set of edges.
    \subsection{Motivation}
    There has been a brief review of constrained Delaunay triangulation algorithms by \cite{Floriani1989} however, since then many algorithms have been reported with each focusing on different aspects of computation like parallel computation, IO efficiency and generalizations like non-Euclidean distance metrics, higher order Delaunay criterion etc. There have also been additions of new algorithms to the categories defined by \cite{Floriani1989} for classification of CDT algorithms. Therefore, objective of this work is to report updates over \cite{Floriani1989}, keeping as much of the taxonomic structure proposed by the original paper as possible and enhancing it wherever it is required. 
    \subsection{Paper outline}
        The concept of Constrained Delaunay triangulation and its properties are introduced in Section 2. Prominent algorithms appearing after the work by \cite{Floriani1989} have been discussed in section 3. Section 4 discusses some interesting generalizations of the CDT problem to higher dimensions, different space metric and a generalization of Constrained Delaunay criteria. Section 5 concludes the paper.   
        
\section{Basic definition and properties of CDT}
\begin{figure}
    \centering
   \begin{subfigure}[b]{0.30\textwidth}
        \centering
        \resizebox{\linewidth}{!}{\begin{tikzpicture}
	    \draw (1, 0) -- (0, 1); 
	    \draw (0, 1) -- (0, -1);
	    \draw (1, 0) -- (0, -1);
	    \draw (0, 0) circle (1cm);
	    \filldraw (1, 0) circle[radius=0.5pt];
	    \filldraw (0, 1) circle[radius=0.5pt];
	    \filldraw (0, -1) circle[radius=0.5pt];
	    \filldraw (-1, 0) circle[radius=0.5pt];
	    \draw (-0.3, 1.5) -- (-0.1, -1.5);
	    \filldraw (-0.3, 1.5) circle[radius=0.5pt];
	    \filldraw (-0.1, -1.5) circle[radius=0.5pt];
	    \node [red, right] at (1, 0) {A};
	    \node [red, above] at (0, 1) {B};					
	    \node [red, left] at (-0.3, 1.5) {C};					
	    \node [red, right] at (-0.1, -1.5) {D};
	    \node [red, right] at (-1, 0) {E};					
	    \node [red, right] at (0, -1) {F};					
        \end{tikzpicture}}
        \caption{Satisfies CDT criterion}
        \label{cdtExplainA}
    \end{subfigure}   
    \begin{subfigure}[b]{0.30\textwidth}
        \centering
        \resizebox{\linewidth}{!}{\begin{tikzpicture}
	    \draw (1, 0) -- (0, 1); 
	    \draw (0, 1) -- (0, -1);
	    \draw (1, 0) -- (0, -1);
	    \draw (0, 0) circle (1cm);
	    \filldraw (1, 0) circle[radius=0.5pt];
	    \filldraw (0, 1) circle[radius=0.5pt];
	    \filldraw (0, -1) circle[radius=0.5pt];
	    \filldraw (0.3, 0.95) circle[radius=0.5pt];
	    \draw (-0.3, 1.5) -- (-0.1, -1.5);
	    \filldraw (-0.3, 1.5) circle[radius=0.5pt];
	    \filldraw (-0.1, -1.5) circle[radius=0.5pt];
	    \node [red, right] at (1, 0) {A};
	    \node [red, above] at (0, 1) {B};					
	    \node [red, left] at (-0.3, 1.5) {C};					
	    \node [red, right] at (-0.1, -1.5) {D};
	    \node [red, right] at (0.3, 0.95) {E};					
	    \node [red, right] at (0, -1) {F};					
        \end{tikzpicture}} 
        \caption{Does not satisfy CDT criterion}
        \label{cdtExplainB}
    \end{subfigure}
\caption{Constrained Delaunay criterion}
\label{cdtExplain}
\end{figure}
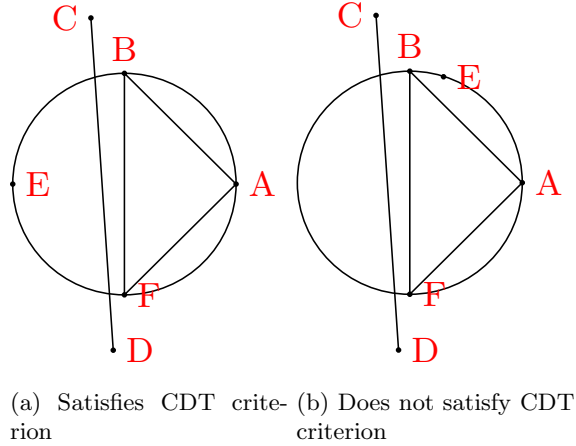

Let \textit{X} be a planar straight line graph(PSLG), then CDT of \textit{X} is the union of all constrained Delaunay simplices, which in two dimensions, contains all segments and points of \textit{X}. A simplex is \textit{constrained Delaunay} if there is a corresponding circumcircle which encloses no point of \textit{X} that is \textit{visible} from the \textit{inside} of the simplex\cite{lee1986generalized}. A point is \textit{visible} from inside a simplex if there is no segment of \textit{X} which intersects a segment drawn between that point and a point in the simplex\cite{Shewchuk:1998:CGE:276884.276893}. As it can be noted, \textit{visibility} criterion relaxes the original \textit{empty-circumcircle} constraint imposed on Delaunay simplices to permit points on or inside the circumcircle of a simplex if they are guarded by a constraint segment against all points of that simplex. Figure \ref{cdtExplain} explains the constrained Delaunay criterion, where in Fig. \ref{cdtExplainA} highlights the case when an outside point(E) is allowed to be at the circumference since it is not visible to any of the constituent points(A, B and C) and Fig. \ref{cdtExplainB} shows the case when an external point(E) is \textit{visible} and hence the triangle ABC is \textit{not} constrained Delaunay.\par

\subsection{Properties of CDT in 2D}
Constrained Delaunay triangulation in two dimensions exhibits identical properties as that by \textit{unconstrained} Delaunay triangulation. For example, like Delaunay triangulation, CDT maximizes the minimum angle of every triangle in the mesh\cite{lee1986generalized}, it minimizes the maximum enclosing circle radius\cite{shewchuk2008general} among all possible constrained triangulations of a given PSLG. Delaunay triangulation and voronoi diagram are duals of each other, similarly, CDTs are duals of constrained(or Bounded) Voronoi diagrams\cite{Klein:1993:LRA:160985.161008}, \cite{Joe1993}, \cite{wang1995finding}.

\section{Proposed taxonomy of state-of-art in CDT algorithms}
Constrained Delaunay triangulation algorithms proposed in the literature can be categorised mainly in two classes based on how they process the input PSLG. They either process it all at once to create the CDT, called the \textit{static} algorithms or they process the input points and constraint segments one at a time thereby imparting incremental nature to the solution, which we categorize as \textit{dynamic} algorithms. 

\subsection{Static CDT algorithms}
Initial work in the direction of development of algorithms for CDT was for simple polygons as inputs. A linear time randomized divide and conquer approach was proposed by \cite{Klein:1993:LRA:160985.161008} for computing \textit{constrained} Voronoi diagram of a simple polygon by merging Voronoi diagrams of sub-regions of the input, thereby using the duality it can be used for computing the corresponding CDT. Specifically, \cite{Klein:1993:LRA:160985.161008} proposed partitioning input polygon into a set of simpler polygons which are called \textit{pseudonormal histograms}(or PNHs). Solution to the original CDT problem is obtained after merging CDTs of these individual PNHs. Decomposition of a polygon into PNHs and merging CDTs of PNHs to make the solution, both stages have linear time complexity however the algorithm for computing CDT of individual PNH was proposed using a random algorithm with linear time expected complexity. \cite{chin1998finding} later improved over this work by proposing a linear-time \textit{deterministic} algorithm for computing CDT of individual PNHs. 

\cite{lee1986generalized} proposed an $O(|n|log |n|)$ average-case time complexity divide and conquer algorithm, where $n$ represents the number of points, for computing CDT(referred to as \textit{Generalized} Delaunay triangulation) of a given simple polygon by merging GDTs of its decomposition into a set of simpler polygons.  \par

\subsubsection{Algorithms for general polygons}
\cite{lee1986generalized} also proposed an $O(|V^2|)$ time CDT algorithm for the case of general graphs(i.e., holes possible), where $|V|$ represents the number of points. Their algorithm is based on identifying the Delaunay edges incident on every point. It starts by first computing a set of points visible from each point and connecting them to form a structure called \textit{visibility graph}, followed by scanning of all edges incident on each point and removing all non-Delaunay edges unless it is a constraint edge. Resultant structure forms the CDT of given set of points and edges. 

\cite{Chew:1987:CDT:41958.41981} proposed an $O(nlog n)$ time static divide and conquer algorithm for computing CDT, it takes complete PSLG as input at once and partitions the space into rectangular strips, computes CDT of individual strips and then combines neighboring strips to compute the CDT of given PSLG. 

\cite{agarwal2005efficient} proposed an IO-efficient CDT algorithm which they claimed experimentally to be able to process 10GBs of LIDAR data on 128MB RAM and within 7.5 hours. There approach was to initially generate recursive subdivisions of the input point set(called, \textit{gradation}), use an already existing internal memory algorithm to compute CDT of the leaf of that recursive subdivision and use this result to compute CDT of supersets progressively(i.e., following the gradation). However, their algorithm is limited to the cases where number of constraint segments of PSLG are in the order of size of main memory. Specifically, for cases where $|S| \geq c_{o}M$, they claim total number of IO operations to be $O(\frac{N}{B}\log_{\frac{M}{B}} {\frac{N}{B}})$. They also pose an open question on existence of a randomized incremental CDT algorithm with $O(nlog_2 n)$ complexity which was later addressed in \cite{shewchuk2015fast}.

\subsection{Dynamic CDT algorithms}
Dynamic CDT algorithms incrementally process input PSLG by inserting constraints one at time in the resultant mesh. Specifically, it inserts input points followed by insertion of constrained edges one at a time as opposed to the static algorithms which process the input all at once. Dynamic algorithms provide the practical flexibility of adding constraints on-demand. Dynamic CDT algorithms proposed thus far in the literature primarily differ in two stages of the algorithm, namely, the strategy for insertion of an input point and that for insertion of a constraint edge in the current triangulation. 

\subsubsection{Point insertion strategy}
\cite{de1988constrained} presents a dynamic algorithm for realizing multi-resolution surface representation using CDT\cite{de1992line} as its basis. It assumes triangulated PSLG as its input. Triangulation of higher resolution surface $T_{i+1}$ is obtained from that of lower resolution, $T_i$ by addition of points and constraint segments. For each new point, it identifies the set of triangles in $T_i$ which will have this point in their circumcircles, called the \textit{influence region} of that point and computes a polygon from outer boundaries of these triangles called \textit{influence polygon}. It then joins new point with all points of influence polygon, thereby re-meshing the interior of influence polygon with addition of this new point. Since only linear number of triangles are affected by insertion of a point, worst-case time complexity of point insertion algorithm is $O(n)$. \cite{Lu:1991:DCD:902954} presents an exactly similar point insertion approach but with different terminology. 

\cite{ANGLADA1997215} proposes a point insertion algorithm derived from generalization of the approach proposed in  \cite{Sloan:1987:FAC:22847.22852},\cite{lawson1977software}. It locates the triangle(say \textit{t}) which encloses the new point, \textit{p}. It then partitions \textit{t} into three triangles \textit{$t_1$},\textit{$t_2$}, \textit{$t_3$}. For each triangle, it determines if its neighbor triangle(sharing non-constraint edge) which does not share \textit{p} does not have \textit{p} in its circumcircle. If this neighbor triangle violates this condition then  it is removed from the mesh and its neighbors are explored in-turn. This process continues until non of the triangles contain \textit{p} in its interior. Worst case complexity of this point insertion procedure is $O(n)$. 

\cite{domiter2004constrained} proposes an approach based on \cite{vzalik2003incremental}. It performs point insertion in three stages, \textit{initialization}, \textit{triangulation} and \textit{finalization}. During initialization, an artificial triangle(also called \textit{super-triangle}) is constructed, it contains encloses all points of input PSLG. This supertriangle is splitted by insertion of first input point into three sub-triangles. Then during triangulation stage, it uniformly subdivides the input region using a two-level uniform planar subdivision data structure(referred as 2LUPS). This data structure partitions the input region in terms of cells, with uniform subdivision called the level one subdivision and adaptive subdivision inside a cell to adapt to non-uniform point density called the level two subdivision. Point insertion proceeds after this subdivision, in which, the point is first inserted into a cell of 2LUPS and then a closest point is searched within that cell. If the triangle incident on this found point also contains the point to be inserted then this triangle is divided else the next closest point is searched and similar checks are repeated. Each sub-triangle is checked for empty circle criterion. Last step in this algorithm is removing all triangles which share a vertex of super-triangle. This algorithm works with $O(n^{1.1})$ average case complexity and in worst case complexity reaches to $O(n^{2})$, where $n$ is the number of inserted points.

\cite{kallmann2004fully} first locates the input point, if it is present on an edge then edge is splited to contain this point, if this point is found on a face then face is splitted. So, point insertion primalrily involves dealing with case of inserting point on an edge or inserting point on a face. Since insertion of a new point may turn an edge non-Delaunay, edge flipping is used to restore Delaunay property of all non-constrained segments. Insertion of one point may require $O(n)$ edge flips, however for Delaunay triangulations with random input expected number of edge flips are constant\cite{guibas1992randomized}.

\subsubsection{Constraint edge insertion strategy}
\cite{de1988constrained}, \cite{de1992line} employ simple generalization of their algorithm for point insertion to the segment insertion problem. For each constraint segment they first identify the list of triangle which intersect this segment which is called the \textit{influence region} of that segment like that for points. From this list, outer boundary of these triangle is identified called the \textit{influence polygon} using an $O(n)$ worst case time algorithm. The new segment \textit{t} to be inserted is a diagonal of this polygon. Therefore, we have two cavities separated by this constrained segment. This polygon is then triangulated(non-Delaunay) using an $O(|Q^2|)$ worst case time algorithm to fill this region, where $|Q|$ is the size of influence polygon. After this re-triangulation, all edges inside this polygon are \textit{optimized} by enforcing the empty-circle criterion.

\cite{sloan1993fast} proposed a similar approach based on incremental insertion of edges which first computes Delaunay triangulation of the input point set using the approach described in \cite{Sloan:1987:FAC:22847.22852} followed by enforcing constraint segments into it and then \textit{optimizing} the triangulation by ensuring that all non-constraint edges in the resultant triangulation(i.e., the CDT) are Delaunay. It loops over every constraint edge,  and for each edge, it finds all intersecting edges in the initial Delaunay triangulation. It then removes all intersecting edges and restores the Delaunay triangulation for non-constraint edges using \textit{triangle swapping}(implemented using edge-flipping). It then removes all superfluous triangles which are either present beyond input boundary or if they share a point with the supertriangle\cite{Sloan:1987:FAC:22847.22852} computed during Delaunay triangulation. They experimentally observe that the proposed algorithms roughly takes CPU time proportional to the number of points(i.e., $O(n)$) in PSLG.

Triangle package employs an algorithm proposed by \cite{shewchuk1996triangle} for computing CDT dynamically. They achieve $O(nlogn)$ average-case time complexity, regardless of distribution of points. It is similar in overall layout to the \cite{sloan1993fast} in that it starts with an initial Delaunay triangulation of input point set, followed by recovery of constraint edges in the final mesh by deleting the triangle it overlaps and \textit{re-triangulating} each side of the region thus formed.

\cite{ANGLADA1997215} proposed an improvement over \cite{de1992line} on constraint edge insertion strategy. It follows the typical outline of incrementally inserting constraint edge, identifying and removing intersecting triangles thereby forming two cavity and then re-triangulating cavities separately. However, the difference from \cite{de1992line} lies in the way they re-triangulate the cavities. They use a recursive algorithm for triangulating upper($P_{u}$) and lower($P_{l}$) polygons(or cavity). In a polygon(say $P_{u}$), this algorithm identifies a point $c$ (say, $c = s_l$) with respect to the constraint edge(say \textit{ab}) such that $\Delta abc$ satisfies empty circumcircle property. $\Delta abc$ divides $P_{u}$ into two subregions $P_E = \{a, s_{1}, s_{2}, ...s_{l}\}$ and $P_D = \{s_l, ..., s_n, b\}$. Then this algorithm is recursively applied on these sub-regions with respect to edges $ac$ and $bc$. This approach ensures preservation of $ab$ in the final mesh after both $P_u$ and $P_l$ are triangulated. Their algorithm has $O(n^2)$ worst-case time complexity. The approach proposed by \cite{shewchuk1996triangle} and \cite{ANGLADA1997215} have established as general framework for constraint segment insertion approaches, \cite{kallmann2004fully}, \cite{domiter2004constrained} and many other works have utilized these algorithms.

\cite{shewchuk2015fast} uses similar segment insertion algorithm framework as in \cite{ANGLADA1997215}, but proposes a new randomized cavity re-triangulation algorithm, along which the resultant segment insertion algorithm has time complexity linear in number of edges crossed by the constraint segment. \cite{shewchuk2015fast} claims that their algorithm can deal with non-convex cavities, dangling segments inside cavity and cavities with self-intersections which was not possible with the CDT algorithm proposed by \cite{Chew:1987:CDT:41958.41981}. They derived a $O(nlogn + nlog^2 k)$ tight bound on average case time complexity using the results from \cite{agarwal2005efficient}. 
 
\subsection{Parallel CDT algorithms for PSLG}
 Parallelization often requires domain decomposition, and in context of computing CDT of input PSLG, we need to partition the space of points in PSLG. In that direction, \cite{chernikov2008algorithm} proposes a quality Delaunay mesh generation algorithm which uses constrained segments to separate sub-domains. Assuming such a domain decomposition, their algorithm utilizes the fact that if each sub-domain has Delaunay conforming mesh then resultant global mesh will also be constrained Delaunay. However, ensuring Delaunay property in each sub-domain may require subdivision of some of its elements, therefore this algorithm may result in subdivision of constrained segments itself in which case a SPLIT message is signaled to the adjacent sub-domain(i.e., a thread or processor) to maintain consistency of the state of constrained segments across sub-domains. Their comparative evaluation on a single node with the algorithm  proposed by \cite{shewchuk1996triangle} suggests comparable running times.

Recently, GPU's have found many applications beyond graphics computing and some of their variants have now come to be referred as General purpose GPUs(or the GPGPU's). \cite{qi2013computing} proposes a CDT algorithm using parallelization advantages available in GPUs. They first construct a triangulation of the points in input PSLG, then constraint edges are inserted using \textit{edge-flipping} which results in a constrained triangulation which is  then transformed into the corresponding Constrained Delaunay triangulation, again using the edge-flipping approach. They use NVIDIA CUDA-enabled GPUs to achieve a claimed speedup of more than 10x over CDT implementation in Triangle package. However, performance of their approach degrades when the input dataset is skewed.

\section{Generalizations}
In context of the usual distance metric for CDT problem, \cite{vigo2000computing} have proposed a generalization to the isotropic nature of triangles generated in conventional CDT algorithms, they propose \textit{directional} CDT, in which if each input point has an associated deformed ellipse which represents curvature of the surface, triangles in the resulting mesh with their shapes adapted to that curvature information can be generated. Such modelling problems arise in mesh generation over parametric surfaces. Their work required generalizing CDT problem from conventional Euclidean distance metric to the elliptical space. Therefore, effectively the circumcircle in the conventional problem becomes a circumellipse in this generalization. Similarly, in this transformed space Delaunay criterion becomes the empty circumellipse criterion.

Addressing the possibility of CDT in higher dimensions, \cite{Shewchuk:1998:CGE:276884.276893} proposes a sufficient condition for the existence of CDT in dimensions higher than two. CDTs have not been generalized beyond two-dimensions because of the existence of many singular non-triangulable structures, for example the Sch\"{o}nhardt polyhedron\cite{schonhardt1928zerlegung}. However, if we can transform our input PSLG to another topologically equivalent input following the Shewchuk's theorem, a CDT exists. It states that an \textit{d-dimensional} input PSLG \textit{X} has CDT if each of its \textit{k-dimensional} constraining facet is a union of k-dimensional \textit{strongly} Delaunay simplices($k \leq d-2$), where, a simplex is strongly Delaunay if there exists a circumsphere for its points which does not contain any other point inside or on its surface(hence a \textit{stronger} version of the Delaunay criterion). 

\cite{gudmundsson2002higher} relaxes the empty circle criterion to include \textit{atmost k} points inside circle, in which case the triangle is called a k-order triangle. Such generalization is useful in optimizing criteria other than characteristic criteria of Delaunay triangulation, for example minimizing the number of local minima(or maxima), which are useful for dealing with artificial dam problem and interrupted ridge lines in elevation models respectively. \cite{gudmundsson2005constrained} further generalizes this concept from Higher order Delaunay triangulation(HODs) to constrained Higher order Delaunay triangulations. In the same direction, \cite{silveira2009towards} proposes definitions of higher order CDT in an attempt to combine the concept of higher order Delaunay triangulation and CDT. These definitions deal with the way of defining the \textit{order} of a triangle. Specifically, it defines different cases for how number of points inside circumcircle of a triangle should be counted, choice of which depends on the application.

\section{Conclusion}
We have observed that in the class of static CDT algorithms, initial work has been in the direction of CDT algorithms for simple polygons and some works have extended the problem domain to general graph(i.e., allowing possibility of holes). Divide and conquer strategy has been a dominant design approach for this class of problems.

Since the review article \cite{Floriani1989}, dynamic CDT algorithms have seen higher growth, which seems to be due to their higher practicability and flexibility as compared to static algorithms which requires complete input all at once. However, since that period the average-case complexity has still stayed around $O(nlogn)$. Almost all reviewed  dynamic CDT algorithms use identical segment insertion algorithm framework but they differ in their strategies for triangulation of polygons(or cavities) created after removal of triangles intersecting the constraint edge. 

Algorithms with focus over IO-efficiency since then have been proposed which are able to handle PSLG's of the size of 10GBs. In present time, such algorithms become very crucial when the data generated by even personal computing devices is of this order. 

Parallel algorithms have been proposed for multi-CPU and recently popular GPGPUs which have facilitated orders of magnitude speedup over equivalent sequential algorithms. 

Further, almost throughout this period, the work by Dr. Shewchuk has formally established properties of constrained Delaunay triangulation \cite{shewchuk2008general}, defined existence criterion for higher dimensions\cite{Shewchuk:1998:CGE:276884.276893} and facilitated a robust 2D CDT code in Triangle\cite{shewchuk1996triangle}. In addition, a recent work of his group on CDT problem \cite{shewchuk2015fast} has attempted answering questions posed by \cite{agarwal2005efficient} regarding existence of randomized incremental CDT algorithm with $O(nlogn)$ time complexity. 

Efforts have also been made to generalize the concept of constrained Delaunay triangulation beyond two dimensions. In order to deal with parametric surface meshes, setting of CDT problem has been extended beyond conventional Euclidean distance metric to deal with elliptical distance metric. Work by \cite{gudmundsson2005constrained} explored generalization of empty circumsphere property of Delaunay elements itself as they have dealt with the case of allowing multiple points inside the circumpsphere and proposed various definitions of Higher order Constrained Delaunay triangulation.

\subsubsection*{Acknowledgements}
Authors would like to thank Computer Division, Bhabha Atomic Research Centre for providing cooperation and support throughout the duration of this work.

\bibliographystyle{splncs03}
\bibliography{2DCDTAlgoSurvey}

\end{document}